\documentclass[10pt,letterpaper]{article}
\usepackage[top=0.85in,left=1in,footskip=0.75in,marginparwidth=2in]{geometry}

\usepackage[utf8]{inputenc}
\usepackage{amsmath,amssymb,bbm}
\usepackage[shortlabels]{enumitem}

\usepackage{float, graphicx}
\usepackage[ruled,vlined,linesnumbered]{algorithm2e}

\usepackage{color}
\usepackage[usenames,dvipsnames]{xcolor}
\definecolor{PurpleViridis}{RGB}{75,0,85}
\definecolor{DarkBlueViridis}{RGB}{0,112,148}
\definecolor{YellowViridis}{RGB}{253,227,51}

\usepackage[aboveskip=1pt,singlelinecheck=off,font=small,labelfont=bf, labelsep=period]{caption}
\usepackage{subcaption}

\usepackage[backend=biber, style=ieee, sorting=ynt, maxcitenames=3,style=alphabetic]{biblatex}
\DeclareDelimFormat{nameyeardelim}{\addcomma\space} 
\usepackage{csquotes}
\usepackage{hyperref, xurl} 
\usepackage{cleveref}

\usepackage{microtype}
\DisableLigatures[f]{encoding = *, family = * }

\raggedright
\usepackage{lastpage,fancyhdr, epstopdf}
\fancyheadoffset[L]{2.25in}
\fancyfootoffset[L]{2.25in}

\newcommand \RR {\mathbb{R}}

\DeclareMathOperator*{\var}{\mathbb{V}\text{ar}}
\DeclareMathOperator*{\cov}{\mathbb{C}\text{ov}}

\newcommand \VV [1]{\var\left[{#1}\right]}
\newcommand \CC [1]{\cov\left[{#1}\right]}

\DeclareMathOperator*{\argmin}{arg\,min}
\DeclareMathOperator*{\diag}{Diag}

\DeclareMathOperator*{\DET}{Det}
\DeclareMathOperator*{\Tr}{Tr}

\newcommand{\CRANpkg}[1]{\href{https://CRAN.R-project.org/package=#1}{\textbf{#1}}}

\newcommand{\code}[1]{\verb|#1|}

\usepackage{awesomebox}
\newcommand{\conclulogo}[1]{\awesomebox{0pt}{\faMarker}{black}{\textbf{Conclusion:} #1}}
\newcommand{\infologo}[1]{\awesomebox{0pt}{\faInfoCircle}{black}{\textbf{Information:} #1}}
\newcommand{\warninglogo}[1]{\awesomebox{0pt}{\faExclamationTriangle}{black}{\textbf{Warning:} #1}}

\usepackage[most]{tcolorbox}
\newtcolorbox{blackbox}[1]{colback=white, colframe=black, 
coltext=black, boxsep=1.5pt, arc=4pt, before=\centering, title=#1}

\newtcolorbox{info}[1]{colback=green!15!white,  coltitle =white, colframe=green,
 coltext=black, boxsep=1.5pt, arc=5pt, before=\centering, 
 title={\infologo{#1}}, drop shadow, boxrule=1mm}
\newtcolorbox{warning}[1]{colback=red!15!white, colframe=red, 
coltext=black, boxsep=1.5pt, arc=5pt, before=\centering,
title={\warninglogo{#1}}, drop shadow, boxrule=1mm}
\newtcolorbox{conclusion}[1]{colback=orange!15!white, colframe=orange, coltext=black,
boxsep=1.5pt, arc=5pt, before=\centering, title={\conclulogo{#1}},
 drop shadow, boxrule=1mm}

\usepackage{multicol} 
 \tcbuselibrary{theorems} 
\tcbset{ 
defstyle/.style={fonttitle=\bfseries\upshape, fontupper=\slshape,
arc=2pt, colback=blue!5!white,colframe=blue!75!black},
theostyle/.style={fonttitle=\bfseries\upshape, fontupper=\slshape,
colback=red!10!white,colframe=red!75!black, arc=2pt, every float=\centering},
propstyle/.style = {fonttitle=\bfseries\upshape, fontupper=\slshape,
 arc=2pt, colback=green!5!white,colframe=green!75!black},
proofstyle/.style = {colback=white,
  colframe=black,  coltext=black, arc=2pt}
}

\newtcbtheorem[number within=section,Crefname={definition}{definitions}]%
{Definition}{Definition}{defstyle}{def}
\newtcbtheorem[use counter from=Definition,Crefname={theorem}{theorems}]%
{Theorem}{Theorem}{theostyle}{th}
\newtcbtheorem[use counter from=Theorem,Crefname={Proof}{Proofs}]%
{Proof}{Proof}{proofstyle}{proof}
\newtcbtheorem[use counter from=Definition,Crefname={corollary}{corollaries}]%
{Corollary}{Corollary}{theostyle}{cr}
\newtcbtheorem[number within=section,Crefname={property}{properties}]%
{Property}{Property}{propstyle}{pr}

\begin{document}
\vspace*{0.35in}

\begin{flushleft}
{\Large
\textbf
\newline{DeCovarT, a multidimensional probalistic model for the deconvolution of heterogeneous transcriptomic samples}
}
\newline
\\
Bastien Chassagnol\textsuperscript{1,2,*},
Grégory Nuel\textsuperscript{2},
Etienne Becht\textsuperscript{1}
\\
\bigskip
\bf{1} Institut De Recherches Internationales Servier (IRIS), FRANCE
\\
\bf{2} LPSM (Laboratoire de Probabilités, Statistiques et Modélisation), Sorbonne Université, 4, place Jussieu, 75252 PARIS, FRANCE
\\
\bigskip
* bastien\_chassagnol@laposte.net

\end{flushleft}

\section*{Abstract}
Although bulk transcriptomic analyses have greatly contributed to a better understanding of complex diseases, their sensibility is hampered by the highly heterogeneous cellular compositions of biological samples. To address this limitation, computational deconvolution methods have been designed to automatically estimate the frequencies of the cellular components that make up tissues,  typically using reference samples of physically purified populations. However, they perform badly at differentiating closely related cell populations. 

We hypothesised that the integration of the covariance matrices of the reference samples could improve the performance of deconvolution algorithms. We therefore developed a new tool, DeCovarT, that integrates the structure of individual cellular transcriptomic network to reconstruct the bulk profile. Specifically, we inferred the ratios of the mixture components by a standard maximum likelihood estimation (MLE) method, using the Levenberg-Marquardt algorithm to recover the maximum from the parametric convolutional distribution of our model.  We then consider a reparametrisation of the log-likelihood to explicitly incorporate the simplex constraint on the ratios.  Preliminary numerical simulations suggest that this new algorithm outperforms previously published methods, particularly when individual cellular transcriptomic profiles strongly overlap. 

\clearpage 

\section{Introduction}
The analysis of the bulk transcriptome provided new insights on the
mechanisms underlying disease development. However, such methods ignore
the intrinsic cellular heterogeneity of complex biological samples, by
averaging measurements over several distinct cell populations. Failure
to account for changes of the cell composition is likely to result in a
loss of \emph{specificity} (genes mistakenly identified as
differentially expressed, while they only reflect an increase in the
cell population naturally producing them) and \emph{sensibility} (genes
expressed by minor cell populations are amenable being masked by highly
variable expression from major cell populations).

Accordingly, a range of computational methods have been developed to
estimate cellular fractions, but they perform poorly in discriminating
cell types displaying high phenotypic proximity. Indeed, most of them
assume that purified cell expression profiles are fixed observations,
omitting the variability and intrinsically interconnected structure of
the transcriptome. For instance, the gold-standard deconvolution algorithm \emph{CIBERSORT} \parencite{newman_etal15} applies nu-support vector regression ($\nu$-SVR) to recover the minimal subset of the most informative genes in the purified signature matrix. However, this machine learning approach assumes that the transcriptomic expressions are independent.

In contrast to these approaches, we hypothesised that
integrating the pairwise covariance of the genes into the reference
transcriptome profiles could enhance the performance of transcriptomic
deconvolution methods. The generative probabilistic model of our algorithm, \emph{DeCovarT} (Deconvolution using the Transcriptomic Covariance), implements this integrated approach.

\section{Model}

First, we introduce the following notations:
\begin{itemize}
    \item $(\boldsymbol{y}=(y_{gi}) \in \mathbb{R}_+^{G\times N}$ is the global bulk transcriptomic expression, measured in $N$ individuals.
    \item  $\boldsymbol{X}=(x_{gj}) \in \mathcal{M}_{\RR^{G\times J}}$ the signature matrix of the mean expression of $G$ genes in $J$ purified cell populations.
    \item $\boldsymbol{p}=(p_{ji})\in ]0, 1[^{J \times N}$ the unknown relative proportions of cell populations in $N$ samples
\end{itemize}

As in most traditional deconvolution models, we assume that the total bulk expression can be reconstructed by summing the individual contributions of each cell population weighted by its frequency, as stated explicitly in the following linear matricial relationship (\Cref{eq:deconvolution-problem}):

\begin{equation}
\label{eq:deconvolution-problem}
\boldsymbol{y}=\boldsymbol{X} \times \boldsymbol{p}
\end{equation}

In addition, we consider unit simplex constraint on the cellular ratios, $\boldsymbol{p}$ (\Cref{eq:positive-ratios}):
\begin{equation}
\label{eq:positive-ratios}
\begin{cases}
\sum_{j=1}^J p_{j}=1\\
\forall j \in \widetilde{J} \quad p_j\ge 0
\end{cases}
\end{equation} 

\subsection{Standard linear deconvolution model}

However, in real conditions with technical and environmental
variability, strict linearity of the deconvolution does not usually hold. Thus, an additional error term is usually considered, and without further assumption on the distribution of this error term, the usual approach to retrieve the best of parameters is by minimising the squared error term between the mixture expressions predicted by the linear model and the actual observed response. This optimisation task is achieved through the ordinary least squares (OLS) approach (\Cref{eq:OLS-task}), 

\begin{equation}
 \label{eq:OLS-task}
 \boldsymbol{\hat{p}}_{i}^{\text{OLS}} \equiv \argmin_{\boldsymbol{p}_i} ||\hat{\boldsymbol{y}_i} - \boldsymbol{y}_i||^2 = \argmin_{\boldsymbol{p}_i}  ||\boldsymbol{X}\boldsymbol{p}_i - \boldsymbol{y}_i||^2 = \sum_{g=1}^G \left( y_{gi} - \sum_{j=1}^J  x_{gj} p_{ji}\right)
\end{equation}

If we additionally assume that the stochastic error term follows a \emph{homoscedastic} zero-centred Gaussian distribution and that the value of the observed covariates (here, the purified expression profiles) is determined (see the corresponding graphical representation in \Cref{subfig:DAG-model-linear} and the set of equations describing it \Cref{eq:white-gaussian-noise}), 
\begin{equation}
    \label{eq:white-gaussian-noise}
     y_{gi} = \sum_{j=1}^J  x_{gj}p_{ji} + \epsilon_i, \quad y_{gi} \sim \mathcal{N} \left(\sum_{j=1}^J  x_{gj}p_{ji}, \sigma_{i}^2 \right), \quad  \epsilon_i \sim  \mathcal{N}(0, \sigma_i^2)
\end{equation}
then, the MLE is equal to the OLS, which, in this framework, is given explicitly by \Cref{eq:OLS-estimate}:
\begin{equation}
    \label{eq:OLS-estimate}
    \boldsymbol{\hat{p}}_{i}^{\text{OLS}} = (\boldsymbol{X}^\top\boldsymbol{X})^{-1}\boldsymbol{X}^\top \boldsymbol{y}_i
\end{equation}

and is known under the the Gauss-Markov theorem.

\subsection{Motivation of using a probabilistic convolution framework}

In contrast to standard linear regression models, we relax in the DeCovarT modelling framework the \emph{exogeneity} assumption, by considering the set of covariates
\(\boldsymbol{X}\) as random variables rather than fixed measures,
in a process close to the approach of DSection algorithm and DeMixt algorithms.
However, to our knowledge, we are the first to weaken the independence assumption between observations by explicitly considering a multivariate distribution and integrating the intrinsic covariance structure of the transcriptome of each purified cell population. 

To do so, we conjecture that the \(G\)-dimensional
vector \(\boldsymbol{x}_j\) characterising the transcriptomic expression
of each cell population follows a multivariate Gaussian distribution, given by \Cref{eq:multivariate-gaussian}:
\begin{equation}
\label{eq:multivariate-gaussian}
    \DET(2\pi\boldsymbol{\Sigma}_j)^{-\frac{1}{2}} \exp\left( -\frac{1}{2} (\boldsymbol{x}_j- \boldsymbol{\mu}_{.j}) \boldsymbol{\Sigma}_j^{-1} (\boldsymbol{x}_j - \boldsymbol{\mu}_{.j})^\top\right)
\end{equation}
and parametrised by:
\begin{itemize}
    \item $\boldsymbol{\mu}_{.j}$, the mean purified transcriptomic expression of cell population $j$
    \item $\boldsymbol{\Sigma}_{j}$, the \textit{positive-definite} (see Definition \Cref{def:positive-definite}) covariance matrix of each cell population. Precisely, we retrieve it from inferring its inverse, known as the precision matrix, through the gLasso \cite{mazumder_hastie11} algorithm.  We define $\boldsymbol{\Theta}_j \equiv \boldsymbol{\Sigma}_j^{-1}$ the corresponding \textit{precision matrix}, whose inputs, after normalisation, store the partial correlation between two genes, conditioned on all the others. Notably, pairwise gene interactions whose corresponding off-diagonal terms in the precision matrix are null are considered statistically spurious, and discarded.  
\end{itemize}

\begin{figure}
     \centering
     \begin{subfigure}{0.5\textwidth}
         \centering
         \caption{Standard linear model representation.}
         \includegraphics[width=\textwidth]{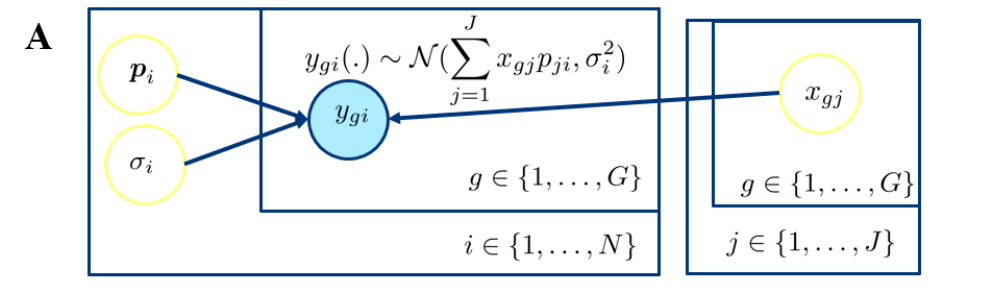}         
         \label{subfig:DAG-model-linear}
     \end{subfigure}
     \begin{subfigure}{0.5\textwidth}
         \centering
         \caption{The generative model used for the DeCovarT framework.}
         \includegraphics[width=\textwidth]{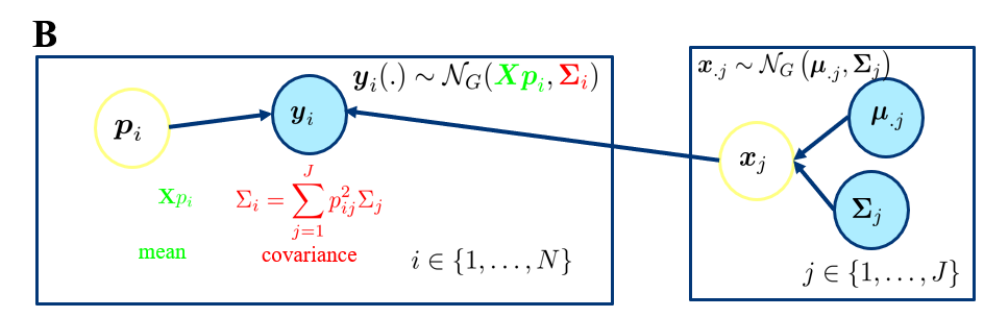}         
         \label{subfig:DAG-model-DeCovarT}
    \end{subfigure}
        \caption[DAG model]{We use the standard graphical convention of graphical models, as depicted in \href{https://revbayes.github.io/tutorials/intro/graph_models.html}{RevBayes} webpage. For identifiability reasons, we conjecture that all variability proceeds from the stochastic nature of the covariates.}
\end{figure}

To derive the log-likelihood of our model, first we \textit{plugged-in} the mean and covariance parameters
\(\zeta_j=\left(\boldsymbol{\mu}_{.j}, \boldsymbol{\Sigma}_j\right)\)
estimated for each cell population in the previous step. 
Then, setting \(\boldsymbol{\zeta}=(\boldsymbol{\mu}, \boldsymbol{\Sigma}), \quad \boldsymbol{\mu}=(\boldsymbol{\mu}_{.j})_{j \in \widetilde{J}} \in \mathcal{M}_{G \times J}, \quad \boldsymbol{\Sigma} \in \mathcal{M}_{G \times G}\) the known parameters and \(\boldsymbol{p}\) the unknown cellular ratios, we show that the conditional distribution of the observed bulk mixture, conditioned on the individual purified expression profiles and their ratios in the sample,
\(\boldsymbol{y}|(\boldsymbol{\zeta}, \boldsymbol{p})\), is the convolution of pairwise independent multivariate Gaussian distributions. 
Using the \textit{affine invariance} property of Gaussian distributions, we can show that this convolution  is also a multivariate Gaussian distribution, given by \Cref{eq:conditional-multivariate-distribution}.
\begin{equation}
\boldsymbol{y}|(\boldsymbol{\zeta}, \boldsymbol{p}) \sim \mathcal{N}_G(\boldsymbol{\mu} \boldsymbol{p}, \boldsymbol{\Sigma}) \text{ with } \boldsymbol{\mu} = (\boldsymbol{\mu}_{.j})_{j \in \widetilde{J}}, \quad \boldsymbol {p}=(p_1, \ldots, p_J) \text{ and } \boldsymbol{\Sigma}=\sum_{j=1}^J p_{j}^2\boldsymbol{\Sigma}_{j}
\label{eq:conditional-multivariate-distribution}
\end{equation}. The DAG associated to this modelling framework is shown in Figure \Cref{subfig:DAG-model-DeCovarT}).

In the next section, we provide an explicit formula of the log-likelihood of our probabilistic framework, its gradient and hessian, which in turn can be used to retrieve the MLE of our distribution. 

\subsection{Derivation of the log-likelihood}

From \Cref{eq:conditional-multivariate-distribution}, the conditional log-likelihood is readily computed and given by \Cref{eq:loglikelihood-multivariate-gaussian}:

\begin{equation}
\ell_{\boldsymbol{y} | \boldsymbol{\zeta}}(\boldsymbol{p})=C + \log\left(\DET \left(\sum_{j=1}^J p_{j}^2\boldsymbol{\Sigma}_{j}\right)^{-1}\right) - \frac{1}{2} (\boldsymbol{y} - \boldsymbol{p} \boldsymbol{\mu})^\top \left(\sum_{j=1}^J p_{j}^2\boldsymbol{\Sigma}_{j}\right)^{-1} (\boldsymbol{y} - \boldsymbol{p}\boldsymbol{\mu})
\label{eq:loglikelihood-multivariate-gaussian}
\end{equation}

\subsection{First and second-order derivation of the unconstrained DeCovarT log-likelihood function}
\label{subsec:uncontrained-optimisation}

The stationary points of a function and notably maxima, are given by the roots (the values at which the function crosses the $x$-axis) of its gradient, in our context, the vector: $\nabla \ell: \RR^J \to \RR^J$ evaluated at point $\nabla \ell (\boldsymbol{p}): ]0, 1[^J \to \RR^J$. Since the computation is the same for any cell ratio $p_j$, we give an explicit formula for only one of them (\Cref{eq:derivative-log-likelihood-unconstrained}):

\begin{equation}
\label{eq:derivative-log-likelihood-unconstrained}
\begin{split}
 \frac{\partial \ell_{\boldsymbol{y} | \boldsymbol{\zeta}}(\boldsymbol{p})}{\partial p_j} =& \scriptstyle \frac{\partial \log\left(\DET(\boldsymbol{\Theta})\right)}{\partial p_j} -\frac{1}{2} \left[\frac{\partial (\boldsymbol{y} - \boldsymbol{\mu} \boldsymbol{p})^\top}{\partial p_j}\boldsymbol{\Theta}(\boldsymbol{y} - \boldsymbol{\mu} \boldsymbol{p}) + (\boldsymbol{y} - \boldsymbol{\mu} \boldsymbol{p})^\top\frac{\partial\boldsymbol{\Theta}}{\partial p_j}(\boldsymbol{y} - \boldsymbol{\mu} \boldsymbol{p}) + (\boldsymbol{y} - \boldsymbol{\mu} \boldsymbol{p})^\top\boldsymbol{\Theta} \frac{\partial (\boldsymbol{y} - \boldsymbol{\mu} \boldsymbol{p})}{\partial p_j} \right]\\
=&  \scriptstyle -\Tr \left(\boldsymbol{\Theta} \frac{\partial \boldsymbol{\Sigma}}{\partial p_j} \right) - \frac{1}{2} \left[ - \boldsymbol{\mu}_{.j}^\top\boldsymbol{\Theta}(\boldsymbol{y} - \boldsymbol{\mu} \boldsymbol{p}) - (\boldsymbol{y} - \boldsymbol{\mu} \boldsymbol{p})^\top\Theta\frac{\partial \Sigma}{\partial p_j}\Theta(\boldsymbol{y} - \boldsymbol{\mu} \boldsymbol{p}) - (\boldsymbol{y} - \boldsymbol{\mu} \boldsymbol{p})^\top\boldsymbol{\Theta}  \boldsymbol{\mu}_{.j} \right] \\
=& \textcolor{PurpleViridis}{-2p_j \Tr \left(\boldsymbol{\Theta}\boldsymbol{\Sigma}_j\right)} +
\textcolor{DarkBlueViridis}{(\boldsymbol{y} - \boldsymbol{\mu} \boldsymbol{p})^\top\boldsymbol{\Theta}  \boldsymbol{\mu}_{.j}} \, +
\textcolor{YellowViridis}{p_j (\boldsymbol{y} - \boldsymbol{\mu} \boldsymbol{p})^\top\boldsymbol{\Theta} \Sigma_j \boldsymbol{\Theta} (\boldsymbol{y} - \boldsymbol{\mu} \boldsymbol{p})}
\end{split}
\end{equation}

Since the solution to $\nabla \left( \ell_{\boldsymbol{y} | \boldsymbol{\zeta}}(\boldsymbol{p}) \right) =0$ is not closed, we had to approximate the MLE using iterated numerical optimisation methods. Some of them, such as the Levenberg–Marquardt algorithm, require a second-order approximation of the function, which needs the computation of the Hessian matrix. Deriving once more \Cref{eq:derivative-log-likelihood-unconstrained} yields the Hessian matrix,  $\mathbf{H} \in \mathcal{M}_{J \times J}$ is given by: 
\begin{equation}
    \label{eq:hessian-log-likelihood-unconstrained}
\begin{aligned}
\mathbf{H}_{i,i}& =
   \frac{\partial^2 \ell}{\partial^2 p_i} =
\textcolor{PurpleViridis}{-2\Tr \left(\boldsymbol{\Theta}\boldsymbol{\Sigma}_i\right) + 4p_i^2 \Tr \left(\left(\boldsymbol{\Theta}\boldsymbol{\Sigma}_i\right)^2\right)}
\textcolor{DarkBlueViridis}{-2p_i(\boldsymbol{y} - \boldsymbol{\mu} \boldsymbol{p})^\top\boldsymbol{\Theta} \boldsymbol{\Sigma}_i \boldsymbol{\Theta} \boldsymbol{\mu_{.i}}\, - \boldsymbol{\mu}_{.i}^\top\boldsymbol{\Theta} \boldsymbol{\mu_{.i}}} \, - \\
& \textcolor{YellowViridis}{2p_i (\boldsymbol{y} - \boldsymbol{\mu} \boldsymbol{p})^\top \boldsymbol{\Theta}\boldsymbol{\Sigma}_i\boldsymbol{\Theta}\boldsymbol{\mu}_{.i} \, -
(\boldsymbol{y} - \boldsymbol{\mu} \boldsymbol{p})^\top\boldsymbol{\Theta} \left(4p_i^2 \boldsymbol{\Sigma}_i \boldsymbol{\Theta} \boldsymbol{\Sigma}_i - \boldsymbol{\Sigma}_i \right)\boldsymbol{\Theta} (\boldsymbol{y} - \boldsymbol{\mu} \boldsymbol{p})}, \quad i \in \widetilde{J} \\
\mathbf{H}_{i,j} &=
   \frac{\partial^2 \ell}{\partial p_i \partial p_j} =
\textcolor{PurpleViridis}{4p_j p_i \Tr \left(\boldsymbol{\Theta}\boldsymbol{\Sigma}_j \boldsymbol{\Theta}\boldsymbol{\Sigma}_i \right)}
\textcolor{DarkBlueViridis}{-2p_i(\boldsymbol{y} - \boldsymbol{\mu} \boldsymbol{p})^\top\boldsymbol{\Theta} \boldsymbol{\Sigma}_i \boldsymbol{\Theta} \boldsymbol{\mu_{.j}} - \boldsymbol{\mu}_{.i}^\top\boldsymbol{\Theta} \boldsymbol{\mu_{.j}}} \, - \\
& \textcolor{YellowViridis}{2p_j (\boldsymbol{y} - \boldsymbol{\mu} \boldsymbol{p})^\top \boldsymbol{\Theta}\boldsymbol{\Sigma}_j\boldsymbol{\Theta} \boldsymbol{\mu}_{.i}\, -
4p_ip_j(\boldsymbol{y} - \boldsymbol{\mu} \boldsymbol{p})^\top\boldsymbol{\Theta}\boldsymbol{\Sigma}_i \boldsymbol{\Theta} \boldsymbol{\Sigma}_j \boldsymbol{\Theta} (\boldsymbol{y} - \boldsymbol{\mu} \boldsymbol{p})}, \quad (i,j) \in \widetilde{J}^2, i \neq j
  \end{aligned}
\end{equation}
in which the coloured sections pair one by one with the corresponding coloured sections of the gradient, given in \Cref{eq:derivative-log-likelihood-unconstrained}. Matrix calculus can largely ease the derivation of complex algebraic expressions, thus we remind in \appendixname~(\textit{Matrix calculus}) relevant matrix properties and derivations \footnote{The numerical consistency of these derivatives was asserted with the \CRANpkg{numDeriv} package, using the more stable Richardson’s extrapolation (\cite{DBLP:journals/toms/Fornberg81}).}.

However, the explicit formulas for the gradient and the hessian matrix of the log-likelihood function, given in \Cref{eq:derivative-log-likelihood-unconstrained} and \Cref{eq:hessian-log-likelihood-unconstrained} respectively, do not take into account the simplex constraint assigned to the ratios. While some optimisation methods use heuristic methods to solve this problem, we consider alternatively a reparametrised version of the problem, detailed comprehensively in Appendix \Cref{subsec:contrained-optimisation}.

\section{Simulations}

\subsection{Simulation of a convolution of multivariate Gaussian mixtures}

To assert numerically the relevance of accounting the correlation between expressed transcripts, we designed a simple toy example with two genes and two cell proportions. Hence, using the simplex constraint (\Cref{eq:positive-ratios}), we only have to estimate one free unconstrained parameter, $\theta_1$, and then uses the mapping function \Cref{eq:mapping-function} to recover the ratios. 

We simulated the bulk mixture, $\boldsymbol{y} \in \mathcal{M}_{G \times N}$, for a set of artificial samples $N=500$, with the following generative model: 

\begin{itemize}
    \item We have tested two levels of cellular ratios, one with equi-balanced proportions ($\boldsymbol{p} = (p_1, p_2=1-p_1)=(\frac{1}{2}, \frac{1}{2})$ and one with highly unbalanced cell populations: $\boldsymbol{p} =(0.95, 0.05)$. 
    \item Then, each purified transcriptomic profile is drawn from a multivariate Gaussian distribution. We compared two scenarios, playing on the mean distance of centroids, respectively $\mu_{.1}=(20, 22), \mu_{.2}=(22, 20)$ and $\mu_{.2}=(20, 40), \mu_{.2}=(40, 20)$) and building the covariance matrix,  $\mathbf{\Sigma} \in \mathcal{M}_{2 \times 2}$ by assuming equal individual variances for each gene  (the diagonal terms of the covariance matrix, $\diag(\boldsymbol{\Sigma_1})=\diag(\boldsymbol{\Sigma_1})=\boldsymbol{I}_2$) but varying the pairwise correlation between gene 1 and gene 2, $\CC{x_{1,2}}$, on the following set of values: $\{-0.8, -0.6, \ldots, 0.8\}$ for each of the cell population.
    \item As stated in \Cref{eq:deconvolution-problem}, we assume that the bulk mixture, $\mathbf{y}_{.i}$ could be directly reconstructed by summing up the individual cellular contributions weighted by their abundance, without additional noise.
\end{itemize}

\subsection{Iterated optimisation}

The extremum, and by extension the MLE, is a root of the gradient of the
log-likelihood. However, in our generative framework, the inverse function cancelling the
gradient of Equation \Cref{eq:loglikelihood-multivariate-gaussian} is non-closed. Instead, iterated numerical optimisation
algorithms that consider first or second-order approximations of the function to optimise are used to approximate the roots.

The \textit{Levenberg-Marquardt (LM)} algorithm  bridges the gap between
between the steepest descent method (first-order) and the Newton-Raphson
method (second-order) by inflating the diagonal terms of the Hessian
matrix. Far from the endpoint, a second-order descent is favoured for
its faster convergence pace, while the steepest approach is privileged
close to the extremum since it allows careful refinement of the step size.
Specially, we used the LM implementation of R package 
\href{https://CRAN.R-project.org/package=marqLevAlg}{\textbf{marqLevAlg}} to infer the ratios $\hat{\boldsymbol{p}}$ from the bootstrap simulations, since it includes an additional convergence criteria, the relative distance to the maximum (RDM), that sets apart extrema from spurious saddle points.

\subsection{Results}
We compared the performance of DeCovarT algorithm with the outcome of a quadratic algorithm that specifically addresses the unit simplex constraint: the negative least squares algorithm (NNLS, \cite{haskell_hanson81}). 

\begin{figure}
    \centering
    \includegraphics[width=0.9\textwidth]{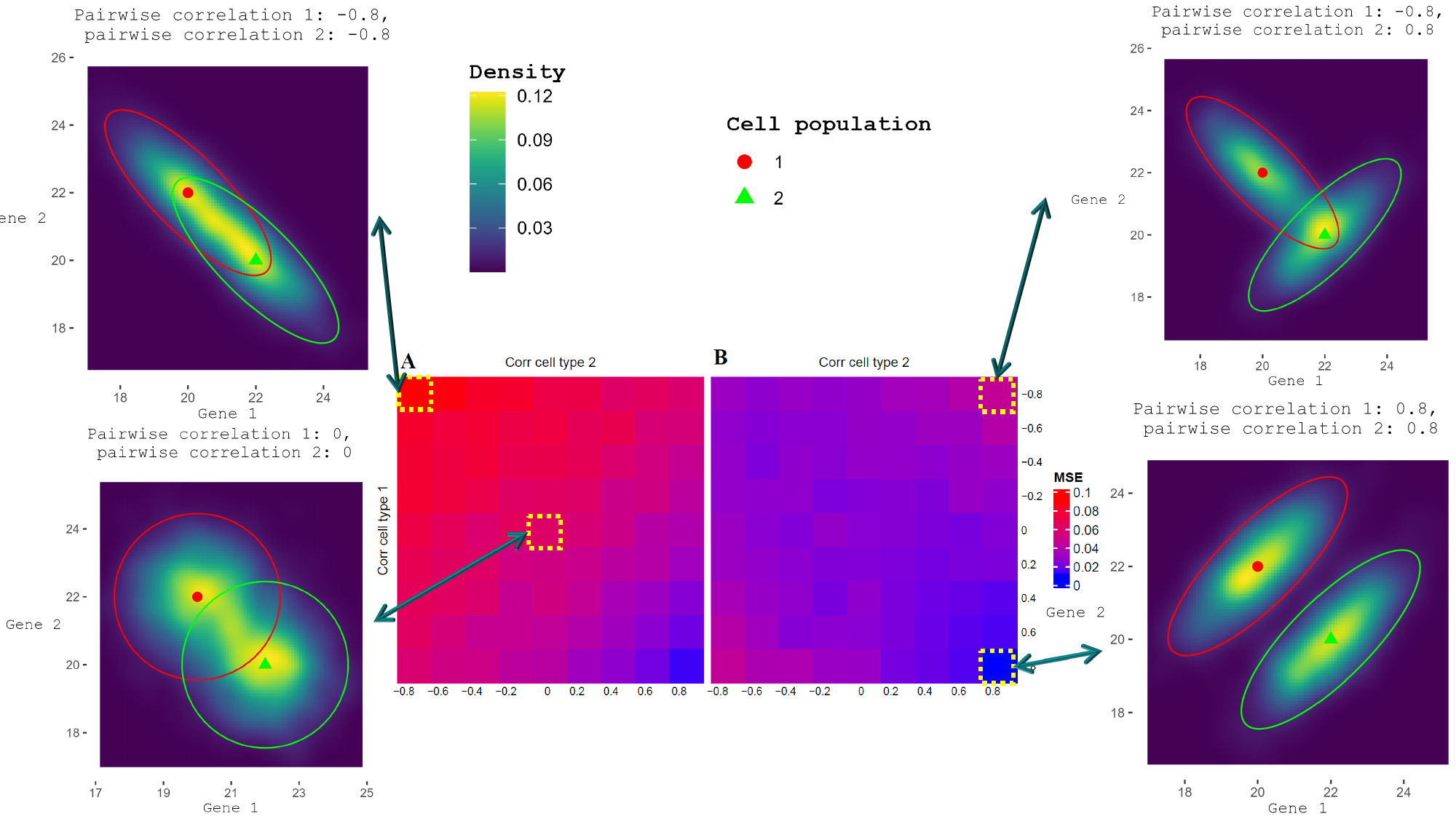}
    \caption{We used the package \CRANpkg{ComplexHeatmap} to display the mean square error (MSE) of the estimated cell ratios, comparing the NNLS output, as implemented in the deconRNASEQ algorithm (\cite{gong_szustakowski13}), in Panel \textbf{A}, with our newly implemented DeCovarT algorithm, in Panel \textbf{B}. The lower the MSE, the least noisy and biased the estimates. In addition, we added the two-dimensional density plot for the intermediate scenario, for which each population is parameterised by a diagonal covariance matrix, and the most extreme scenarios (those with the highest correlation between genes). The ellipsoids represent for each cell population the $95\%$ confidence region and the red spherical icon and the green triangular icon represent respectively the centroids (average expression of gene 1 and gene 2) of cell population 1 and cell population 2.}
    \label{fig:mse-complex_heatmap}
\end{figure}

Even with a limited toy example including two cell populations characterised only by two genes, we observe that the overlap was a good proxy of the quality of the estimation: the less the overlap between the two cell distributions, the better the quality of the estimation \Cref{fig:mse-complex_heatmap}.

The package used to generate the simulations and infer ratios from virtual or real biological mixtures with the DeCovarT algorithm is implemented on my personal Github account \href{https://github.com/bastienchassagnol/DeCovarT}{DeCovarT}.

\section{Perspectives}
The new deconvolution algorithm that we implemented, DeCovarT, is the first one based on a multivariate generative model while complying explicitly the simplex constraint. Hence, it provides a strong basis to further derive statistical tests to assert whether the proportion of a given cell population differs significantly between two distinct biological conditions.

However, we still need to assert its performance in an extended simulation framework. In a numerical setting, we could first increase the dimensionality of our purified datasets by using more realistic parametrisations, using the mean and sparse covariance parameters inferred from purified cellular datasets. Then, we need to evaluate our algorithm in a real-world experience, with both blood and tumoral samples. The Kassandra project would be a good place to start, since the purified database collects a compendium of 9,404 cellular transcriptomic profiles, annotated into 38 blood cellular populations and the performance of Kassandra's algorithm was benchmarked in $N=517$ samples in 6 public datasets with both flow cytometry annotations and bulk RNA-seq expression, against 8 different standard deconvolution algorithms: 5 reference profile deconvolution algorithms: EPIC \cite{racle_etal17}, CIBERSORT \cite{newman_etal15}, CIBERSORTx \cite{newman_etal19}, quanTIseq \cite{finotello_etal19} and ABIS \cite{monaco_etal19}, and 3 marker-based deconvolution algorithms \footnote{Contrary to algorithms based on signature references, marker-based algorithms make the strong asssumption that any discriminant gene, referred to as \textit{marker} is uniquely expresssed in a cell population.}: MCPcounter \cite{becht_etal16} and xCell \cite{aran_etal17}.

Finally, the gLasso algorithm used to derive each purified cell accuracy matrix, like any penalty regularisation approach, is subject to \textit{parameter shrinkage}. Notably, in our setting, shrinkage leads to systematically underestimate the non-zero partial correlations of the precision matrix. A way to circumvent this problem is to only use the \textit{support} (the non-null inputs) output of the gLasso and use the associated topological constraints within a standard MLE approach to fine-tune the inputs of the precision matrix. One way of doing so would be to infer a directed Gaussian Graphical Model (GGM), however, except in really specific topological configurations, such as chordal graphs, there is no current direct equivalence between the space of undirected Markov graphs, as returned by gLasso, and directed Bayesian graphs (\cite{dahl_etal05}).

\printbibliography 

\appendix
\setcounter{figure}{0}
\renewcommand{\thefigure}{S\arabic{figure}}

\section{Optimisation and calculus}

\subsection{Multivariate distributions and basic algebra properties}

\begin{Definition}{Multivariate Gaussian distributions}{multivariate-gaussian}
If random vector $\boldsymbol{X}$ of size $G$ follows a random multivariate Gaussian distribution, $\boldsymbol{X} \sim \mathcal{N}_G(\mu, \boldsymbol{\Sigma})$, then its distribution is given by:
\begin{equation*}
    \DET(2\pi\Sigma)^{-\frac{1}{2}} \exp\left( -\frac{1}{2} (\boldsymbol{x} - \mu) \Sigma^{-1} (\boldsymbol{x} - \mu)^\top\right)
\end{equation*}
in which:
\begin{itemize}
    \item $\mu=\boldsymbol{X}$ is the $G$-dimensional mean vector
    \item $\boldsymbol{\Sigma}$ is a $G\times G$ positive-definite \Cref{def:positive-definite} covariance matrix, whose diagonal terms, $\diag(\boldsymbol{\Sigma})=[(\VV{X_{i,j}}), \, \forall (i, j) \in \widetilde{G}^2, i= j]^\top$ are the individual variances of each purified gene transcript in population $j$ and off-diagonal terms, $\boldsymbol{\Sigma}_{i,j}=\CC{X_{i}, X_{j}}, \, \forall (i, j) \in \widetilde{G}^2, i \neq j$ are the covariance between variables. We note $\Theta=\Sigma^{-1}$, the inverse of the covariance matrix, called the \textit{precision matrix}.
\end{itemize}

\end{Definition}

\begin{Property}{Affine invariance property of multivariate GMMs}{multivariate-affine-transformation}

The two following properties hold for a multivariate Gaussian distribution:
\begin{itemize}
    \item if $\boldsymbol{X} \sim \mathcal{N}_G( \mu, \boldsymbol{\Sigma})$, then $p\boldsymbol{X}$, with $p$ a constant, follows itself a multivariate Gaussian distribution, given by:
    $p\boldsymbol{X} \sim \mathcal{N}_G (p \mu, p^2 \Sigma)$
    
    \item given two independent random vectors $\boldsymbol{X_1} \sim \mathcal{N}_G(\mu_1, \boldsymbol{\Sigma_1})$ and  $\boldsymbol{X_2} \sim \mathcal{N}_G(\mu_2, \boldsymbol{\Sigma_2})$ following a multivariate Gaussian distribution, then the random variable $\boldsymbol{X_1} + \boldsymbol{X_2}$ follows itself the multivariate Gaussian distribution:
    \begin{equation*}
        X + Y \sim \mathcal{N}_G (\mu_1 + \mu_2, \boldsymbol{\Sigma_1} + \boldsymbol{\Sigma_2})
    \end{equation*}
    
    By induction, this property generalises to the sum of $J$ independent random vectors of same dimension $\RR^G$.
\end{itemize}
\end{Property}

Deriving the characteristic function of the multivariate GMM yields directly results reported in \Cref{pr:multivariate-affine-transformation}.

\begin{Definition}{Definite matrix}{positive-definite}
A symmetric real matrix $\boldsymbol{A}$ of rank $G$ is \textit{positive-definite} if:
\begin{equation}
    \label{eq:positive-definite}
    \boldsymbol{x}^\top \boldsymbol{A} \boldsymbol{x} > 0, \quad \boldsymbol{x} \in \mathbb{R}^G
\end{equation}

To gain a clearer grasp of the positive-definite constraint imposed on the covariance parameter of a multivariate Gaussian distribution, let's delve into the most straightforward scenario, in which we assume that any of the individual features exhibit pairwise independence. This particular setup is parametrised by a covariance matrix containing exclusively diagonal elements.

If the matrix is not strictly positive-definite, then some of the diagonal elements can display negative values, otherwise that the individual variances for some of the covariates are negative. It is not physically possible and leads to improper, degenerate probability distributions.
\end{Definition}

\clearpage
\subsection{Matrix and linear algebra}
\label{subsec:linear-algebra}

\begin{Property}{Determinant and trace}{det-trace}
For a squared matrix $A$ of rank $G$ with defined inverse variance $A^{-1}$ and a constant $p$, the following properties hold:
\begin{multicols}{3}
\begin{enumerate}[label=(\alph*)]
\item $\DET(p\boldsymbol{A})=p^G \DET (\boldsymbol{A})$
\columnbreak
\item $\Tr\left(p \boldsymbol{A}\right)=p \Tr(\boldsymbol{A})$
\columnbreak
\item $\DET(A^{-1})=\frac{1}{\DET(A)}$
\end{enumerate}
\end{multicols}    

The trace operator is additionally invariant under \textit{cyclic permutation}, illustrated in \Cref{eq:tr-invariant} for three matrices with matching dimensions: 

\begin{equation*}    
\label{eq:tr-invariant}
\Tr(\boldsymbol{A}\boldsymbol{B}\boldsymbol{C})=\Tr(\boldsymbol{C}\boldsymbol{A}\boldsymbol{B})=\Tr(\boldsymbol{B}\boldsymbol{C}\boldsymbol{A})
\end{equation*}
\end{Property}

\begin{Property}{Transpose}{transpose}
Given two matrices $\boldsymbol{A}$ and $\boldsymbol{B}$, the following properties hold when computing their transpose:
\begin{multicols}{3}
\begin{enumerate}[label=(\alph*)]
    \item $(\boldsymbol{A}^\top)^\top=\boldsymbol{A}$
    \item $(\boldsymbol{A}\boldsymbol{B})^\top=\boldsymbol{B}^\top\boldsymbol{A}^\top$
    \item $\left(\boldsymbol{A}^{-1}\right)^\top=\boldsymbol{A}^{-1}$\footnotetext{with $\boldsymbol{A}$ a symmetric matrix.} 
\end{enumerate}
\end{multicols}

Given two vectors $\boldsymbol{x}$ and $\boldsymbol{y}$ in $\RR^G$ and $\boldsymbol{A}$ a symmetric matrix of rank $G$, using the properties described above, we have \Cref{eq:transpose-vector}
\begin{equation}
    \label{eq:transpose-vector}
    \boldsymbol{x}^\top \boldsymbol{A} \boldsymbol{y} = \boldsymbol{y}^\top \boldsymbol{A} \boldsymbol{x}
\end{equation}
\end{Property}

\clearpage

\subsection{Matrix calculus}
\label{subsec:matrix-calculus}

Fundamental algebra calculus formulas used to derive first-order (\Cref{eq:derivative-log-likelihood-unconstrained}) and second-order (\Cref{eq:hessian-log-likelihood-unconstrained}) derivates are reported in \Cref{pr:matrix-calculus-first-order} and \Cref{pr:matrix-calculus-second-order}, respectively.

\begin{Property}{First-order matrix calculus}{matrix-calculus-first-order}

Given two invertible matrices,  $A=\boldsymbol{A}(p)$ and $B=\boldsymbol{B}(p)$, functions of a scalar variable $p$, the following matrix calculus hold:

\begin{multicols}{3}
\begin{enumerate}[label=(\alph*)]
\item $\frac{\partial \DET(\boldsymbol{A})}{\partial p}=\DET(\boldsymbol{A}) \Tr \left(\boldsymbol{A}^{-1} \frac{\partial \boldsymbol{A}}{\partial p} \right)$
\item $\frac{\partial \boldsymbol{U}\boldsymbol{A}\boldsymbol{V}}{\partial p} = \boldsymbol{U} \frac{\partial\boldsymbol{A}}{\partial p} \boldsymbol{V}$
\item $\frac{\partial \boldsymbol{A}^{-1}}{\partial p} = -\boldsymbol{A}^{-1} \frac{\partial\boldsymbol{A}}{\partial p} \boldsymbol{A}^{-1}$
\end{enumerate}
\end{multicols}

From a) and fundamental linear algebra properties enumerated in \Cref{subsec:linear-algebra}, we can readily compute applying the chain rule property on the logarithm:

\begin{equation*}
    \begin{split}
\frac{\partial \log\left(\DET(\boldsymbol{A})\right)}{\partial p}&= \Tr \left(\boldsymbol{A}^{-1} \frac{\partial \boldsymbol{A}}{\partial p} \right) \\
    \frac{\partial \log\left(\DET(\boldsymbol{A}^{-1})\right)}{\partial p}&=- \Tr \left(\boldsymbol{A}^{-1} \frac{\partial \boldsymbol{A}}{\partial p} \right)
\end{split}
\end{equation*}

Finally, injecting these first-order matrix derivatives with \Cref{pr:transpose} we have:

\begin{equation*}
\begin{split}
\frac{\partial (\boldsymbol{y} - \boldsymbol{x} p)^\top \Theta (\boldsymbol{y} - \boldsymbol{x} p)}{\partial p} & = -2  (\boldsymbol{y} - \boldsymbol{x} p)^\top \Theta \boldsymbol{x} \\
&= -2 \boldsymbol{x}^\top \Theta (\boldsymbol{y} - \boldsymbol{x} p) \\
\text{ with } \boldsymbol{A} = \boldsymbol{D} = - \boldsymbol{x} \in \RR^G, \quad 
\boldsymbol{b} = &\boldsymbol{e} = \boldsymbol{y}, \quad
\boldsymbol{C}= \boldsymbol{\Theta} \text{ symmetric}
\end{split}
\end{equation*}
\end{Property}




\begin{Property}{Second-order matrix calculus}{matrix-calculus-second-order}
Given an invertible matrix $\boldsymbol{A}$ depending on a variable $p$, the following calculus formulas hold:

\begin{enumerate}[label=(\alph*)]
\begin{minipage}[c]{0.7\linewidth}
\item $  \frac{\partial^2 \boldsymbol{A}^{-1}}{\partial p_i \partial p_j} =  \boldsymbol{A}^{-1} \left( \frac{\partial \boldsymbol{A}}{\partial p_i } \boldsymbol{A}^{-1} \frac{\partial \boldsymbol{A}}{\partial p_j} - \frac{\partial^2 \boldsymbol{A}}{\partial p_i \partial p_j}  + \frac{\partial \boldsymbol{A}}{\partial p_j } \boldsymbol{A}^{-1} \frac{\partial \boldsymbol{A}}{\partial p_i}\right) \boldsymbol{A}^{-1} \quad
$
\end{minipage} 
\begin{minipage}[c]{0.25\linewidth}
\item $
 \quad \frac{\partial \Tr \left(\boldsymbol{A}\right)}{\partial p_i} = \Tr  \left(\frac{\partial \boldsymbol{A}}{\partial p_i}\right)
$
\end{minipage}
\end{enumerate}

Combining \Cref{pr:matrix-calculus-first-order} with the linear property of the trace operator yields:

\begin{equation*}
    \frac{\partial^2 \log \left(\DET (\boldsymbol{A}^{-1})\right)}{\partial^2 p} = - \Tr  \left[\boldsymbol{A}^{-1}\frac{\partial^2 \boldsymbol{A}}{\partial^2 p_i}\right] +
\Tr  \left[\left(\boldsymbol{A}^{-1}\frac{\partial \boldsymbol{A}}{\partial p_i}\right)^2\right]
\end{equation*}

\end{Property}

\clearpage

\subsection{First and second-order derivation of the constrained DeCovarT log-likelihood function}
\label{subsec:contrained-optimisation}

To reparametrise the log-likelihood function (\Cref{eq:loglikelihood-multivariate-gaussian}) in order to explicitly handling the unit simplex constraint (\Cref{eq:positive-ratios}), we consider the following mapping function: $\boldsymbol{\psi}:\boldsymbol{\theta}  \to  \boldsymbol{p} \, | \quad  \boldsymbol{\theta} \in \RR^{J-1} , \, \boldsymbol{p} \in ]0, 1[^{J}$ (\Cref{eq:mapping-function}):

\begin{multicols}{2}
\begin{enumerate}
\item \begin{equation}
\label{eq:mapping-function}
\boldsymbol{p} = \boldsymbol{\psi} (\boldsymbol{\theta}) =
\begin{cases}
p_j =  \frac{e^{\theta_j}}{\sum_{k < J} e^{\theta_k} \, + \, 1}, \, j < J\\
p_J =  \frac{1}{\sum_{k < J} e^{\theta_j} + 1}
\end{cases}
\end{equation}
\item $\boldsymbol{\theta} = \boldsymbol{\psi}^{-1} (\boldsymbol{p}) = \left(\ln{\left( \frac{p_j}{p_J}\right)} \right)_{j \in \{ 1, \ldots, J -1\}}$
\end{enumerate}
\end{multicols} that is a $C^2$-diffeomorphism, since $\boldsymbol{\psi}$ is a bijection between $\boldsymbol{p}$ and $\boldsymbol{\theta}$ twice differentiable. 

Its Jacobian, $\mathbf{J}_{\boldsymbol{\psi}} \in \mathcal{M}_{J \times (J-1)}$ is given by \Cref{eq:mapping-function-gradient}: 

\begin{equation}
\label{eq:mapping-function-gradient} 
\mathbf{J}_{i,j} =
   \frac{\partial p_i}{\partial \theta_{j}} =
\begin{cases}
\frac{e^{\theta_i}B_i}{A^2 },\quad i = j, \, i < J\\
\frac{-e^{\theta_j}e^{\theta_i}}{A^2 }, \quad i \neq j, \, i < J\\
\frac{-e^{\theta_j}}{A^2}, \quad i=J
\end{cases}
\end{equation}

with $i$ indexing vector-valued $\boldsymbol{p}$ and $j$ indexing the first-order order partial derivatives of the mapping function, $A=\sum_{j' < J} \,e^{\theta_{j'}} \, +  \, 1$ the sum over exponential (denominator of the mapping function) and $B=A - e^{\theta_{i}}$ the sum over ratios minus the exponential indexed with the currently considered index $i$. 

The Hessian of the multi-dimensional mapping function $\boldsymbol{\psi (\theta)}$ exhibits symmetry for each cell ratio component $j$, as anticipated in accordance with Schwarz's theorem. It is is a third-order tensor of rank $(J-1)(J-1)J$, given by \Cref{eq:mapping-function-hessian}:

\begin{equation}
    \label{eq:mapping-function-hessian}
    \begin{aligned}
   \frac{\partial^2 p_i}{\partial k \partial j} &=
\begin{cases}
\frac{e^{\theta_i} e^{\theta_l} \left (-B_i + e^{\theta_i}\right)}{A^3},\, (i<J) \land \left((i\neq j) \oplus(i\neq k)\right) \quad (a)\\
\frac{2 e^{\theta_i} e^{\theta_j} e^{\theta_k} }{A^3}, \, (i<J) \land  (i \neq j \neq k)  \quad (b)\\
\frac{e^{\theta_i} e^{\theta_j} \left (-A + 2e^{\theta_j}\right)}{A^3}, \, (i<J) \land (j=k\neq i)  \quad (c)\\
\frac{B_i e^{\theta_i} \left( B_i -  e^{\theta_i}\right)}{A^3}, \, (i<J) \land (j = k = i)  \quad (d)\\
\frac{e^{\theta_j} \left( -A + 2 e^{\theta_j}\right)}{A^3}, \, (i=J) \land (j = k)  \quad (e)\\
\frac{2 e^{\theta_j} e^{\theta_k}}{A^3}, \, (i=J) \land (j \neq k)  \quad (f)
\end{cases}
  \end{aligned}
\end{equation}

with $i$ indexing $\boldsymbol{p}$, $j$ and $k$ respectively indexing the first-order and second-order partial derivatives of the mapping function with respect to $\boldsymbol{\theta}$. In line $(a)$, $\oplus$ refers to the Boolean XOR operator, $\land$ to the AND operator and $l=\{j,k\} \setminus i$.

To derive the log-likelihood function in \Cref{eq:derivative-log-likelihood-unconstrained}, we reparametrise $\boldsymbol{p}$ to  $\boldsymbol{\theta}$, using a standard \textit{chain rule formula}. Considering the original log-likelihood function, \Cref{eq:loglikelihood-multivariate-gaussian}, and the mapping function, \Cref{eq:mapping-function}, the differential at the first order and at the second order is given by \Cref{eq:chain-rule-first-order} and \Cref{eq:chain-rule-second-order}, respectively defined in $\RR^{J-1}$ and $\mathcal{M}_{(J-1)\times(J-1)}$:

\begin{equation}
\label{eq:chain-rule-first-order}
\begin{bmatrix}
\frac{\partial \ell_{\boldsymbol{y} | \boldsymbol{\zeta}}}{\partial \theta_j}
\end{bmatrix}_{j < J}
  =  \sum_{i=1}^J \frac{\partial \ell_{\boldsymbol{y} | \boldsymbol{\zeta}}}{\partial p_i} \frac{\partial p_i}{\partial \theta_j} 
\end{equation}

\begin{equation}
\label{eq:chain-rule-second-order}
 \begin{bmatrix}   
   \frac{\partial \ell_{\boldsymbol{y}|\boldsymbol{\zeta}}^2 }{\partial \theta_k \theta_j} 
  \end{bmatrix}_{j < J, \, k < J}   = 
  \sum_{i=1}^J \sum_{l=1}^J \left(  \frac{\partial p_i }{\partial \theta_j} \frac{\partial^2 \ell_{\boldsymbol{y}|\boldsymbol{\zeta}} }{\partial p_i \partial p_l} \frac{\partial p_l }{\partial \theta_k}\right)  \, + \,  \sum_{i=1}^J  \left( \frac{\partial \ell_{\boldsymbol{y}|\boldsymbol{\zeta}} }{\partial p_i} \frac{\partial^2 p_i }{\partial \theta_k \theta_j}\right) \quad (d) 
\end{equation}

\end{document}